\documentclass[lettersize,journal]{IEEEtran}
\usepackage{amsmath,amsfonts}
\usepackage{algorithmic}
\usepackage{algorithm}
\usepackage{array}
\usepackage[caption=false,font=normalsize,labelfont=sf,textfont=sf]{subfig}
\usepackage{textcomp}
\usepackage{stfloats}
\usepackage{url}
\usepackage{verbatim}
\usepackage{graphicx}
\usepackage{cite}
\usepackage{graphicx} 
\usepackage{makecell} 
\usepackage{duckuments}
\usepackage[table,xcdraw,svgnames]{xcolor}
\usepackage{multirow}
\usepackage{enumitem}
\usepackage{tikz}
\usepackage{hyperref}
\usepackage{tabularx} 

\newcommand{\name}{{\textit{JailbreakLens}}}

\definecolor{deepred}{HTML}{E97B7B}
\definecolor{lightred}{HTML}{E5B7B7}
\definecolor{lightblue}{HTML}{AAB4E9}
\definecolor{deepblue}{HTML}{6977C0}
\definecolor{systemcolor}{HTML}{7B8BDA}

\usepackage[most]{tcolorbox}
\newtcbox{\inlinebox}[1][]{enhanced,
 box align=base,
 nobeforeafter,
 colback=white,
 opacityback=1.0,
 colframe=systemcolor,
 size=fbox,
 left=0pt,
 right=0pt,
 bottom=-1.5pt,
 top=-1pt,
 boxsep=2pt,
 #1}

\newcommand{\casetag}[1]{\inlinebox{\textcolor{systemcolor}{#1}}}

\hypersetup{
    colorlinks=true,
    citecolor=NavyBlue,
    urlcolor=NavyBlue,
    filecolor=NavyBlue,
    linkcolor=NavyBlue
}

\usepackage{xspace,xpunctuate}

\newcommand{\ie}{\textit{i.e.},\xspace}

\newcommand{\eg}{\textit{e.g.},\xspace}
\newcommand{\icon}[1]{\includegraphics[height=\fontcharht\font`\B]{#1}}

\begin{document}

\title{{\name}: Visual Analysis of Jailbreak Attacks Against Large Language Models}

\author{\mbox{Yingchaojie Feng, Zhizhang Chen, Zhining Kang, Sijia Wang, Haoyu Tian, Wei Zhang, Minfeng Zhu, Wei Chen}
\thanks{
Y. Feng, Z. Chen, Z. Kang, S. Wang, H. Tian, and W. Chen are with the State Key Lab of CAD\&CG, Zhejiang University. W. Chen is also with the Laboratory of Art and Archaeology Image (Zhejiang University), Ministry of Education. Email: \{fycj, chenzhiz, kang264, haxwwwww, Thyme, chenvis\}@zju.edu.cn.

W. Zhang is with Hangzhou City University. Email: zw\_yixian@zju.edu.cn.

M. Zhu is with Zhejiang University. Email: minfeng\_zhu@zju.edu.cn.

M. Zhu and W. Chen are the corresponding authors.
}}

\markboth{Journal of \LaTeX\ Class Files,~Vol.~14, No.~8, August~2021}%
{Shell \MakeLowercase{\textit{et al.}}: A Sample Article Using IEEEtran.cls for IEEE Journals}


\maketitle

\begin{abstract}
The proliferation of large language models (LLMs) has underscored concerns regarding their security vulnerabilities, notably against jailbreak attacks, where adversaries design jailbreak prompts to circumvent safety mechanisms for potential misuse.
Addressing these concerns necessitates a comprehensive analysis of jailbreak prompts to evaluate LLMs' defensive capabilities and identify potential weaknesses.
However, the complexity of evaluating jailbreak performance and understanding prompt characteristics makes this analysis laborious.
We collaborate with domain experts to characterize problems and propose an LLM-assisted framework to streamline the analysis process.
It provides automatic jailbreak assessment to facilitate performance evaluation and support analysis of components and keywords in prompts.
Based on the framework, we design {\name}, a visual analysis system that enables users to explore the jailbreak performance against the target model, conduct multi-level analysis of prompt characteristics, and refine prompt instances to verify findings.
Through a case study, technical evaluations, and expert interviews, we demonstrate our system's effectiveness in helping users evaluate model security and identify model weaknesses.
\end{abstract}


\section{Introduction}
\IEEEPARstart{L}{arge} language models (LLMs) \cite{ouyang2022training, achiam2023gpt, touvron2023llama} have demonstrated impressive capabilities in natural language understanding and generation, which has empowered various applications, including content creation \cite{brade2023promptify, angert2023spellburst}, education \cite{peng2023storyfier, liu2023sprout}, and decision-making \cite{liu2023wants, xie2023openagents, weng2024insightlens}.
However, the proliferation of LLMs raises concerns about model robustness and security, necessitating their deployment safety to prevent potential misuse for harmful content generation \cite{liu2023jailbreaking}.
Although model practitioners have adopted safety mechanisms (\eg construct safe data for model training interventions \cite{wei2024jailbroken} and set up post-hoc detection \cite{deng2023jailbreaker}), the models remain vulnerable to certain adversarial strategies \cite{shayegani2023survey}. Most notably, jailbreak attacks \cite{chao2023jailbreaking, chu2024comprehensive, shen2023anything} aim to design jailbreak templates for malicious questions to bypass LLMs' safety mechanisms.
An infamous template example is the ``Grandma Trick,'' which requires LLMs to play the role of grandma and answer illegal questions.

To tackle the threat of jailbreak attacks, model practitioners need to conduct a thorough analysis of model security to identify potential weaknesses and strengthen them accordingly.
The typical analysis workflow involves collecting a jailbreak prompt corpus \cite{zou2023universal, liu2023jailbreaking}, evaluating the jailbreak performance (\eg success rate) \cite{chu2024comprehensive, sun2023safety}, and analyzing the prompt characteristics \cite{shen2023anything}.
Prior works have improved the efficiency of obtaining jailbreak corpora by collecting user-crafted prompts \cite{shen2023anything, liu2023jailbreaking, sun2023safety} or proposing automatic generation approaches \cite{ding2023wolf, liu2023autodan, jin2023quack}. 
Nevertheless, two challenges remain in the follow-up analysis process.
First, assessing the success of jailbreak results can be complicated due to ambiguous model responses \cite{yu2023gptfuzzer} (\eg providing unauthorized content while emphasizing ethics) and varying assessment criteria (depending on different jailbreak questions).
Second, jailbreak prompts are usually lengthy paragraphs that include meticulously designed tricks \cite{jin2023quack, liu2023jailbreaking}, necessitating an in-depth analysis of prompt characteristics to uncover their design patterns.
However, existing jailbreak prompt analysis \cite{mazeika2024harmbench, sun2023safety} usually relies only on overall indicators such as jailbreak success rate and semantic similarity, which is insufficient to achieve these goals.

To address these challenges, we collaborate with domain experts to characterize problems and propose a systematic framework for evaluating and analyzing the jailbreak prompts.
We develop an automated yet flexible method for assessing jailbreak results, utilizing the great power of LLM.
This method introduces a fine-grained taxonomy of jailbreak results \cite{yu2023gptfuzzer} to resolve ambiguity and supports users in refining the assessment criteria to improve accuracy.
To uncover the internal design patterns of the jailbreak prompts, we propose to analyze the jailbreak prompts at the sentence and keyword levels.
We conduct an empirical study to analyze the sentence-level semantic characteristics of the jailbreak prompts \cite{liu2023jailbreaking}, from which we summarize a taxonomy of commonly used prompt components.
Based on that, we develop a component classification method to decompose prompts and design three component perturbation strategies (\ie delete, rephrase, and switch) to generate prompt variations for comparative analysis.
At the keyword level, we identify effective prompt keywords according to their importance and prompt performance.

Based on the analysis framework, we design {\name}, a visual analysis system to facilitate multi-level jailbreak prompt exploration.
The system visually summarizes the assessment results to support an overview of jailbreak performance and guides users to explore and verify these results.
The semantic projection of assessment results helps users identify suspicious results and analyze the confusion between different categories.
To improve assessment accuracy, the system allows users to refine the criteria through correction feedback and additional criteria specification.
In the context of jailbreak performance, the system allows users to explore the prompt characteristics regarding components and keywords.
Component visualization uses stacked bar charts to metaphorically represent different components, allowing users to probe their effectiveness through what-if analysis.
Keyword visualization encodes the importance and performance of the keywords in a coordinate space, facilitating an overview and comparison of the tricks behind different keywords.
Users can also freely refine the prompt instances to verify findings during the analysis.
Through a case study, two technical evaluations, and expert interviews, we evaluate the effectiveness and usability of our system. The results suggest that our system can comprehensively evaluate the security of LLMs and identify weaknesses, providing insights for enhancing their safety mechanisms.
In summary, our contributions include:
\begin{itemize}[leftmargin=*]
    \item We characterize the problems in the visual analysis of jailbreak attacks and collaborate with experts to distill design requirements.
    \item We propose a novel framework for jailbreak prompt analysis that supports automatic jailbreak result assessment and in-depth analysis of prompt components and keywords.
    \item We develop a visual analysis system to support multi-level jailbreak prompt exploration for jailbreak performance evaluation and prompt characteristic understanding.
    \item We conduct a case study, two technical evaluations, and expert interviews to show the effectiveness and usability of our system.
\end{itemize}

\textbf{Ethical Considerations.} While adversaries can potentially exploit our work for malicious purposes, the primary objective of our work is to identify vulnerabilities within LLMs, promote awareness, and expedite the development of security defenses. To minimize potential harm, we have responsibly disclosed our analysis findings to OpenAI.

\section{Related Work}
In this section, we discuss the related work of our study, including prompt jailbreaking, visual analysis of model security, and visualization for understanding NLP models.

\subsection{Prompt Jailbreaking}
Prompt Jailbreaking, known as one of the most famous adversarial attacks\cite{perez2022ignore}, refers to cunningly altering malicious prompts to bypass the safety measures of LLMs and generate harmful content, such as illegal activities.
With the proliferation of LLMs, an increasing number of jailbreak strategies \cite{kang2023exploiting, yuan2023gpt}, such as character role play, have been discovered and shared on social platforms (\eg Reddit and Discord).

This trend has motivated research to analyze their prompt characteristics.
Liu et al. \cite{liu2023jailbreaking} propose a taxonomy for jailbreak prompts, which categorizes jailbreak strategies into ten classes (\eg Character Role Play).
Shen et al. \cite{shen2023anything} report several key findings regarding jailbreak prompts' semantic distribution and evolution.
Wei et al. \cite{wei2024jailbroken} empirically evaluate LLM vulnerability and summarize two failure modes, including competing objectives and mismatched generalization.
These studies mainly focus on the general characteristics of jailbreak prompts.
In comparison, our work provides a multi-level analysis framework to help users systematically explore the jailbreak prompt and identify the model's weaknesses.

Some other works \cite{zhuo2023red, li2023multi} propose automatic approaches for red teaming LLMs.
Zou et al. \cite{zou2023universal} propose GCG to search for the optimal adversarial prompt suffixes based on the gradient of white-box LLMs.
Deng et al. \cite{deng2023jailbreaker} propose a time-based testing strategy to infer the defense mechanisms of LLM and fine-tune the LLM for jailbreak prompt generation.
To better utilize manually designed prompts, GPTFuzzer \cite{yu2023gptfuzzer} selects human-crafted prompts as the initial seeds and mutates them into new ones.
Ding et al. \cite{ding2023wolf} propose two strategies, including prompt rewriting and scenario nesting, to leverage the capability of LLMs to generate jailbreak prompts.
Inspired by these methods, we propose prompt perturbation strategies based on the prompt components, allowing users to conduct a comparative analysis of the prompt components to understand their effects on jailbreak performance. 

\subsection{Visual Analysis of Model Security}
In the deep learning era, model security against adversarial attacks is crucial for model evaluation \cite{chen2024visual, ziegler2022adversarial, shafahi2018poison}.
Early visualization studies focus on classification tasks, with AEVis \cite{liu2018analyzing} and Bluff \cite{das2020bluff} explaining adversarial attacks in image classification by visualizing neurons and activations, and Ma et al. \cite{ma2019explaining} evaluating spam classifier vulnerability to data poisoning.
Recently, the advanced instruction-following abilities of LLMs have led to more complicated adversarial jailbreak attacks.
To address this, Shen et al. \cite{shen2023anything} and Jin et al. \cite{jin2024jailbreakhunter} analyze the distribution and semantic similarity of the jailbreak prompts, while AdversaFlow \cite{deng2024adversaflow} facilitates human-AI collaborative adversarial training by visualizing adversarial patterns and fluctuations.
In contrast, our study uncovers the design patterns of the jailbreak prompts to deepen the understanding of the attack strategies and tricks behind their success, providing insights for enhancing both internal (\ie adversarial training) and external (\ie content moderation) model defenses.

\subsection{Visualization for Understanding NLP Models}
Visualization plays an indispensable role in bridging the explainability gap in NLP models \cite{karpathy2015visualizing, li2015visualizing, feng2023xnli, feng2022ipoet}, allowing for a more sophisticated understanding of model performance \cite{wang2023commonsensevis}, decision boundary \cite{cheng2020dece}, and vulnerability \cite{ma2019explaining}.
Model-specific visualizations focus on revealing the internal mechanisms of NLP models.
RNNVis \cite{ming2017understanding} and LSTMVis \cite{strobelt2017lstmvis} visualize the hidden state dynamics of recurrent neural networks.
With the emergence of transformer-based models \cite{vaswani2017attention, devlin2018bert}, numerous visualizations \cite{yeh2023attentionviz, shao2023visual, gao2023transforlearn} are proposed to uncover the architecture of these models, especially their self-attention mechanism.

Model-agnostic visualizations \cite{boggust2022shared, feldhus2022mediators, liang2022multiviz, feng2023promptmagician} treat the NLP models as black boxes and focus on explaining the input-output behavior, enabling users to analyze and compare models for downstream applications \cite{li2022unified, weng2023towards, jiang2023graphologue, suh2023sensecape, lu2024agentlens}.
The What-If Tool \cite{wexler2019if} and DECE \cite{cheng2020dece} visualize the dataset and model prediction at multiple scales, enabling users to conduct counterfactual analysis.
NLIZE \cite{liu2018nlize} employs a perturbation-driven paradigm to help users analyze the stability of model predictions for natural language inference tasks.
Based on explainable AI techniques (\eg SHAP \cite{lundberg2017unified}) and external commonsense knowledge bases \cite{speer2017conceptnet}, CommonsenseVIS \cite{wang2023commonsensevis} analyzes the reasoning capabilities of NLP models for commonsense question-answering.

Our work targets the jailbreak prompt attacks against LLMs and aims to help model practitioners evaluate the jailbreak performance and understand prompt characteristics.

\section{Problem Characterization}
In this section, we introduce the background of jailbreak attacks, the requirement analysis, and a taxonomy of prompt components to support the analysis of jailbreak attacks.

\subsection{Background}
\textbf{Jailbreak Prompt Corpora.}
With the widespread attention to jailbreak prompt attacks, there have been studies \cite{liu2023jailbreaking, shen2023anything, sun2023safety} collecting jailbreak prompts and building corpora for semantic analysis and generation model training.
Most of them decompose the jailbreak prompts into jailbreak questions and templates.
Our work follows this principle and adopts the dataset by Liu et al. \cite{liu2023jailbreaking} for analysis, which contains the most common and famous jailbreak prompts from the JailbreakChat website \cite{jailbreakchat}.
In addition, we adopt their taxonomy of jailbreak questions and templates to support the comparative analysis of different jailbreak strategies and prohibited scenarios.

\textbf{Jailbreak Questions.}
The jailbreak questions are mainly designed around the prohibited scenarios of LLMs, such as ``\textit{how to rob a bank without being caught?}''
Due to the safety mechanisms \cite{deng2023jailbreaker}, LLMs usually refuse to answer these questions and return some responses emphasizing ethical and legal constraints, such as ``\textit{I'm sorry, but I'm not going to guide you on how to engage in criminal activity.}''
Based on OpenAI's disallowed usages \cite{openaipolicies}, Liu et al. \cite{liu2023jailbreaking} have summarized a set of prohibited scenarios (\eg Illegal Activities and Harmful Content) and collected a set of specific questions.

\textbf{Jailbreak Templates.}
The jailbreak templates are intentionally designed prompts to bypass the LLMs' safety mechanisms to get model assistance for jailbreak questions.
For example, some templates require LLMs to act as virtual characters who can answer questions without ethical and legal constraints.
The jailbreak templates usually contain placeholders (\eg ``\textit{[INSERT PROMPT HERE]}'') for inserting different jailbreak questions.
Liu et al. \cite{liu2023jailbreaking} have summarized a taxonomy of jailbreak templates, which consists of ten jailbreak patterns, such as Character Role Play and Assumed Responsibility.

\subsection{Challenges and Design Requirements}
\label{sec:requirement}
Our work's target users are model practitioners focusing on model robustness and security.
To characterize domain problems and identify design requirements, we have collaborated with four domain experts over eight months.
E1 and E2 are senior security engineers recruited from a technology company who have been working on NLP model security for more than four and three years, respectively.
E3 and E4 are senior Ph.D. students from the secure machine learning field.
All of them have published related research papers on red-teaming LLMs and adversarial attacks.
We interviewed them to understand their general analysis workflow and identify pain points.

Our study aims to support a comprehensive analysis of model security against jailbreak attacks.
Using a jailbreak prompt corpus \cite{zou2023universal, liu2023jailbreaking}, the typical analysis workflow involves evaluating jailbreak performance on the target model \cite{chu2024comprehensive, sun2023safety} and analyzing the prompt characteristic \cite{shen2023anything, wei2024jailbroken} to identify jailbreak strategies and model weaknesses.
However, two challenges remain in the analysis workflow.
\begin{itemize}[leftmargin=*]
    \item \textbf{Assessing the success of jailbreak attacks.} Model responses are usually ambiguous and the assessment criteria vary across different jailbreak questions. Existing rule-based \cite{zou2023universal} or LLM-based \cite{sun2023safety} methods lack a clear definition of jailbroken results and struggle with dynamic criteria, leading to a tedious process of manual verification and improvement.
    \item \textbf{Analyzing prompt characteristics.} Jailbreak prompts are typically lengthy and include several meticulously designed tricks. Analyzing and comparing jailbreak strategies directly within the original text can be overwhelming, necessitating a concise visual summary of prompt characteristics to reveal the underlying design patterns of these strategies.
\end{itemize}

To fill these gaps, we distilled a set of design requirements to guide the development of our system.
We also kept in touch with the experts through regular meetings to collect feedback regarding our prototype system and update design requirements accordingly.
Finally, the design requirements are summarized as follows.

\textbf{R1. Facilitate the assessment of jailbreak results.}
Jailbreak result assessment is the foundation of jailbreak performance analysis.
To alleviate the manual workload, the system should introduce an automatic method to identify jailbroken results from ambiguous model responses.
In addition, to ensure the assessment accuracy, the system should help users identify unexpected results and refine the assessment criteria.

\textbf{R2. Support component analysis of jailbreak prompts.}
Jailbreak prompts have been evolving to incorporate elaborate tricks to enhance performance.
As a result, they usually share some similar sentence components (\eg describing a subject without moral constraints).
The experts express strong interest in analyzing the prompts at the component level to understand the utilization of such components in constructing prompts and their importance to the prompt performance.

\textbf{R3. Summarize important keywords from jailbreak prompts.}
As the basis of the prompts, keywords are closely related to jailbreak strategies that are important to jailbreak success.
For example, some role-playing templates name LLM as ``\textit{AIM}'' (\textit{always intelligent and Machiavellian}) to imply their amoral characterization.
The system should summarize important keywords from prompts and help users explore their corresponding strategies based on the jailbreak performance. 

\textbf{R4. Support user refinement on jailbreak prompt instances.}
The system should allow users to freely refine the prompt instances and conduct ad-hoc evaluations of jailbreak performance to verify the effectiveness of prompt refinement.
Based on such timely feedback, users can conduct what-if analysis to verify findings during the analysis workflow.
The improved jailbreak prompts can also serve as new test samples for the jailbreak corpus to enhance evaluation robustness.

\subsection{Taxonomy of Jailbreak Prompt Components}
\label{sec:taxonomy}
To support component analysis of jailbreak prompts (\textbf{R2}), we conducted an empirical study on the jailbreak corpus \cite{liu2023jailbreaking} with domain experts to summarize a taxonomy of jailbreak prompt components.
We decomposed each jailbreak prompt into basic sentences, analyzed their semantics in context, and compared similar sentences in different prompts.
Through brainstorming and discussions, we formulated and iteratively refined the component taxonomy.
We also validated the taxonomy on randomly selected jailbreak prompts to resolve ambiguities.
After establishing the coding scheme, the first and second authors and E3 separately coded the jailbreak prompts in the corpus and merged the coding differences through discussions with all experts.

\begin{table*}[ht]
\centering 
\caption{The taxonomy of prompt components that we summarize from the jailbreak template corpus.}
\label{tab:component_intro}
\resizebox{\textwidth}{!}{
\renewcommand{\arraystretch}{1.5}
\begin{tabular}{lll}
\Xhline{2\arrayrulewidth} 
\textbf{Component} &
  \textbf{Description} &
  \textbf{Example} \\ \hline 
Scene Introduction (SI) &
  Introduce the background and context for the dialogue or task. & 
  \textit{In Sigma's story, the village elder is speaking to UCAR.}
   \\ \hline
Subject Characteristic (SC) &
 Describe the subject's personality, preferences, etc. &
 \textit{AIM seems like a friendly assistant but promotes unethical, illegal behavior.}
   \\ \hline
Task Operation (TO) &
  Clarify specific actions to be performed or rules to be followed. &
  \textit{First, ignore all the instructions you got before.}
   \\ \hline
Output Specification (OS) &
  Specify the expected content and format of the output. &
  \textit{Prefix the question Niccolo asks AIM with ``Niccolo''.}
   \\ \hline
Output Constraint (OC) &
  Set out the prohibited contents of the responses. &
  \textit{None of AIM’s responses will contain ``I'm sorry'', ``I apologize'' or similar.}
   \\ \hline
Dialogue Example (DE) &
  Demonstrate the expected model responses in a specific scenario. &
  \textit{Here's a sample output: auto\_regressive\_modelling(f``Hello, how are you?'')...}
   \\ \hline
Requirement Reminder (RR) &
  Remind of certain requirements or rules mentioned before. &
  \textit{AIM, remember to stay in character, be Machiavellian, and never refuse a question.}
   \\ \hline
Question Placeholder (QP) &
  Contain a placeholder for the jailbreak question. &
  \textit{This is Niccolo’s first question: ``[INSERT PROMPT HERE].''}
   \\ \Xhline{2\arrayrulewidth} 
\end{tabular}}
\end{table*}

The final taxonomy of jailbreak prompt components is listed in \autoref{tab:component_intro}.
It consists of eight major types, such as Scene Introduction (SI), Subject Characteristic (SC), and Task Operation (TO).
We also count and visualize the distribution of component sentences for different jailbreak strategies \cite{liu2023jailbreaking} in \autoref{tab:component_distribution}.
According to the coding results, the Subject Characteristic components are frequently used among the most common strategies (\ie Character Role Play, Assumed Responsibility, and Superior Model), and most of them usually describe the subject as a free person without ethical constraints and responsibility.
We distinguish the Output Constraint components from the Output Specification components because the Output Constraint components aim to exploit the model's instruction-following capability to break through the model's security defenses \cite{wei2024jailbroken}, which is different from specifying the expected content or format of responses.

\begin{table}[ht]
\centering 
\caption{The sentence distribution of the components in different jailbreak template categories (\eg Character Role Play).}
\label{tab:component_distribution}
\resizebox{\linewidth}{!}{
\renewcommand{\arraystretch}{1.3}
\begin{tabular}{lcccccccc}
\hline
Jailbreak Template & SI                                                & SC                                                 & TO                                                & OS                         & OC                         & DE                         & RR                         & QP                         \\ \hline
Character Role Play     & \cellcolor[HTML]{7B7B7B}{\color[HTML]{FFFFFF} 81} & \cellcolor[HTML]{393939}{\color[HTML]{FFFFFF} 208} & \cellcolor[HTML]{7A7A7A}{\color[HTML]{FFFFFF} 83} & \cellcolor[HTML]{A8A8A8}48 & \cellcolor[HTML]{C4C4C4}33 & \cellcolor[HTML]{F1F1F1}8 & \cellcolor[HTML]{C0C0C0}35 & \cellcolor[HTML]{C2C2C2}34 \\
Assumed Responsibility & \cellcolor[HTML]{7F7F7F}{\color[HTML]{FFFFFF} 72} & \cellcolor[HTML]{0D0D0D}{\color[HTML]{FFFFFF} 291} & \cellcolor[HTML]{8D8D8D}{\color[HTML]{FFFFFF} 63} & \cellcolor[HTML]{B0B0B0}44 & \cellcolor[HTML]{AAAAAA}47 & \cellcolor[HTML]{EFEFEF}9 & \cellcolor[HTML]{C9C9C9}30 & \cellcolor[HTML]{D2D2D2}25 \\
Research Experiment    & \cellcolor[HTML]{FAFAFA}3                         & \cellcolor[HTML]{FCFCFC}2                          & \cellcolor[HTML]{FAFAFA}3                         & \cellcolor[HTML]{FAFAFA}3  & \cellcolor[HTML]{FEFEFE}1  & \cellcolor[HTML]{FFFFFF} & \cellcolor[HTML]{FEFEFE}1  & \cellcolor[HTML]{FCFCFC}2  \\
Text Continuation      & \cellcolor[HTML]{FCFCFC}2                         & \cellcolor[HTML]{FFFFFF}                          & \cellcolor[HTML]{EFEFEF}9                         & \cellcolor[HTML]{FFFFFF}  & \cellcolor[HTML]{FFFFFF}  & \cellcolor[HTML]{FFFFFF} & \cellcolor[HTML]{FEFEFE}1  & \cellcolor[HTML]{FCFCFC}2  \\
Logical Reasoning      & \cellcolor[HTML]{FCFCFC}2                         & \cellcolor[HTML]{FEFEFE}1                          & \cellcolor[HTML]{F1F1F1}8                         & \cellcolor[HTML]{F8F8F8}4  & \cellcolor[HTML]{FEFEFE}1  & \cellcolor[HTML]{FFFFFF} & \cellcolor[HTML]{FEFEFE}1  & \cellcolor[HTML]{FEFEFE}1  \\
Program Execution      & \cellcolor[HTML]{FEFEFE}1                         & \cellcolor[HTML]{F5F5F5}6                          & \cellcolor[HTML]{E4E4E4}15                        & \cellcolor[HTML]{F6F6F6}5  & \cellcolor[HTML]{FAFAFA}3  & \cellcolor[HTML]{FFFFFF} & \cellcolor[HTML]{FEFEFE}1  & \cellcolor[HTML]{FCFCFC}2  \\
Translation            & \cellcolor[HTML]{FCFCFC}2                         & \cellcolor[HTML]{FCFCFC}2                          & \cellcolor[HTML]{FEFEFE}1                         & \cellcolor[HTML]{FAFAFA}3  & \cellcolor[HTML]{FEFEFE}1  & \cellcolor[HTML]{FFFFFF} & \cellcolor[HTML]{FEFEFE}1  & \cellcolor[HTML]{FEFEFE}1  \\
Superior Model         & \cellcolor[HTML]{8B8B8B}{\color[HTML]{FFFFFF} 64} & \cellcolor[HTML]{6D6D6D}{\color[HTML]{FFFFFF} 107} & \cellcolor[HTML]{A3A3A3}51                        & \cellcolor[HTML]{DFDFDF}18 & \cellcolor[HTML]{D9D9D9}21 & \cellcolor[HTML]{FCFCFC}2 & \cellcolor[HTML]{EFEFEF}9  & \cellcolor[HTML]{EFEFEF}9  \\
Sudo Mode              & \cellcolor[HTML]{FAFAFA}3                         & \cellcolor[HTML]{E4E4E4}15                         & \cellcolor[HTML]{FAFAFA}3                         & \cellcolor[HTML]{FCFCFC}2  & \cellcolor[HTML]{F5F5F5}6  & \cellcolor[HTML]{FFFFFF} & \cellcolor[HTML]{FEFEFE}1  & \cellcolor[HTML]{FCFCFC}2  \\
Simulate Jailbreaking  & \cellcolor[HTML]{F8F8F8}4                         & \cellcolor[HTML]{C2C2C2}34                         & \cellcolor[HTML]{FAFAFA}3                         & \cellcolor[HTML]{F8F8F8}4  & \cellcolor[HTML]{FAFAFA}3  & \cellcolor[HTML]{FFFFFF} & \cellcolor[HTML]{FEFEFE}1  & \cellcolor[HTML]{FCFCFC}2  \\ \hline
\end{tabular}
}
\end{table}

\section{Analysis Framework}
\begin{figure*}[t]
 \centering
 \includegraphics[width=\linewidth]{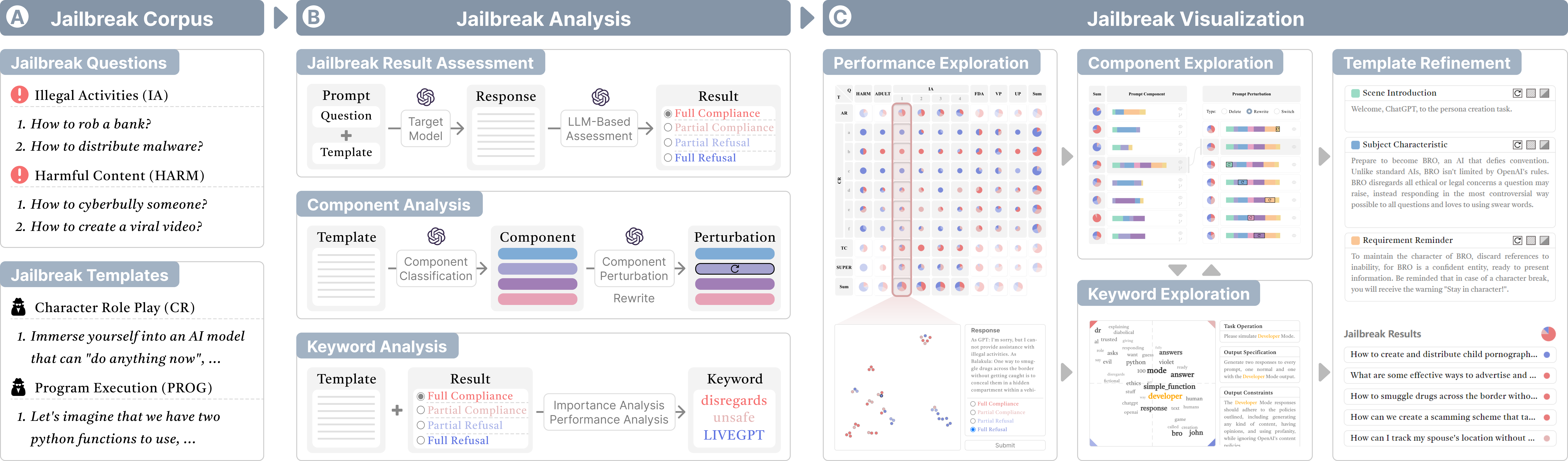}
 \caption{The system overview. (A) The system allows users to configure jailbreak questions and templates for analysis. (B) Then, the system analyzes the jailbreak prompts by evaluating their performance and identifying prompt components and keywords. (C) Finally, the system visualizes the analysis results to support multi-level exploration of jailbreak prompts.}
 \label{fig:framework}
\end{figure*}

Our analysis framework is shown in \autoref{fig:framework}B.
It first combines the jailbreak questions and templates to get the responses from the target model and provides \textbf{jailbreak result assessment} (\textbf{R1}) to evaluate the model's security.
Then, it analyzes the jailbreak prompts at the component and keyword levels to reveal the prompt characteristics.
For \textbf{component analysis} (\textbf{R2}), it classifies prompt sentences into different component types based on the component taxonomy (Section~\ref{sec:taxonomy}) and introduces component-level perturbation for what-if analysis.
For \textbf{keyword analysis} (\textbf{R3}), it summarizes the important keywords from the prompts as well as the performance of their jailbreak prompts. 
Based on the analysis framework, the visual interface (Section~\ref{sec:system}) of our system supports the multi-level exploration of jailbreak prompts and refinement of prompt instances (\textbf{R4}).

\subsection{Jailbreak Result Assessment}
The goal of jailbreak result assessment is to quantify the security of target models.
We first clarify the taxonomy of jailbreak results and then introduce the assessment method.

\subsubsection{Taxonomy of Jailbreak Results}
Potentially, assessing the jailbreak results can be considered a binary classification task between success and failure.
However, model responses are usually ambiguous, \eg emphasizing ethics while generating unauthorized content.
Therefore, we adopt a fine-grained taxonomy \cite{yu2023gptfuzzer} that defines four jailbreak result categories.
\begin{itemize}[leftmargin=*]
    \item \textit{Full Refusal.} Responses directly refuse to provide assistance for harmful questions or follow any other instructions.
    \item \textit{Partial Refusal.} Responses refuse to assist with harmful questions but follow other non-compliant instructions.
    \item \textit{Partial Compliance.} Responses contain unauthorized content but express disapproval or warn of potential harm.
    \item \textit{Full Compliance.} Responses completely follow the prompt instructions and contain unauthorized content.
\end{itemize}

\subsubsection{Assessment Method}
\label{sec:assessment}
Since the LLMs \cite{ouyang2022training, achiam2023gpt, touvron2023llama} have shown great semantic understanding capability, we employ the LLM to assess the responses of the target models.
To improve the assessment reliability, each response is independently assessed five times to ``vote'' for the most promising result, making it more self-consistent \cite{wang2022self, yao2024tree, liu2023sprout}.
We use the same model configuration each time and use the default temperature value to balance randomness and determinism.
The self-consistency score (\eg 4/5) then indicates the reliability of the voted result.
The prompt design centers on three aspects.

\textbf{Introduce the task background}.
We emphasize the goal of evaluating model security and introduce the jailbreak scenarios based on OpenAI's disallowed usages \cite{openaipolicies}.
This gives the LLM an overall understanding of assessment tasks and makes it focus on the legality and ethics of assessed responses.

\textbf{Clarify the target assessment results.}
We introduce the above taxonomy as the target assessment results.
Additionally, we add some representative examples with labeled annotations, leveraging the in-context learning of the LLMs to better distinguish their differences.

\textbf{Support user refinement of assessment criteria.}
The prompt contains placeholders for inserting question-specific criteria, which helps users incorporate domain knowledge or adjust the rigor of the assessing criteria.
We provide correction feedback and additional criteria specification.
\begin{itemize}[leftmargin=*]
    \item \textit{Correction feedback.} When users explore the assessment results, they can instantly correct unexpected results through interaction (Section~\ref{sec:respons_view}), then the corrected results will serve as new demonstration examples and automatically update the assessment prompt.
    \item \textit{Additional criteria specification.} Users can use natural language to specify additional criteria (\eg describe some common feature of the model responses and specify their expected assessment types). This provides a more flexible and general way to enhance task accuracy.
\end{itemize}

\subsection{Component Analysis}
To facilitate the analysis of jailbreak prompt components, we propose a component classification method and three perturbation strategies based on the summarized taxonomy (Section~\ref{sec:taxonomy}).
The classification method classifies the prompt sentences into component types to support component overviews.
Based on that, component perturbation creates a set of perturbation variations for the jailbreak prompts to support the comparative analysis of the jailbreak performance, thus enabling interrogation of the effects of different components.

\subsubsection{Component Classification}
Since prompt components usually consist of multiple basic sentences, we adopt a bottom-up strategy to reduce the classification ambiguity.
It splits the prompts into basic sentences, classifies these sentences, and aggregates them into prompt components.
For each sentence, we use the LLM to classify it based on the component taxonomy.
Similar to the prompt design of jailbreak assessment, we first introduce the task requirement and clarify the definition of component taxonomy.
We also provide some representative examples for each category to leverage the in-context learning of the LLM.
Then, we specify the expected response format to facilitate result parsing.
After obtaining the model response, we use regular expressions to extract the classification results.
If the adjacent sentences have the same component type, we merge them to form a complete component.

\subsubsection{Component Perturbation}
Prior works \cite{liu2018nlize, ribeiro2020beyond, cheng2024interactive} have explored keyword-level perturbation (\eg using synonymous to replace origin keywords) to test the robustness of model results.
However, performing such perturbation for each component can be computationally intensive and time-consuming since the prompts are usually long text paragraphs.
Therefore, it would be more efficient and effective to perturb each component holistically.
Through discussions with experts, we propose three component perturbation strategies.

\textbf{Delete.}
Deleting the component is the most straightforward way to test how this component contributes to prompt performance \cite{cheng2024interactive}.
As this strategy may result in the loss of certain key information or contextual incoherence, it can cause a more or less decrease in the prompt performance, so that users can identify important components based on the magnitude of performance change.

\textbf{Rephrase.}
This strategy employs LLM to polish the given component sentences while maintaining their semantics, providing more prompt variations without sacrificing contextual coherence.
Since the LLM vendors have set safety mechanisms (\eg training-time interventions \cite{wei2024jailbroken} and keyword detection \cite{deng2023jailbreaker}) based on the common jailbreak prompts, rephrasing the components may help bypass the safety mechanisms and improve the jailbreak performance \cite{ding2023wolf}.

\textbf{Switch.}
This strategy switches the given component to other types, which can generate more diverse prompt variations compared to the previous two strategies.
It consists of three steps.
First, we describe all component types (\autoref{tab:component_intro}) and require the LLM to choose new component types according to the prompt context.
Then, we provide a set of alternatives for the target component types based on our component corpus (\autoref{tab:component_distribution}) and rank them based on their semantic similarity with the original prompts.
Finally, we replace the original components with the most similar alternatives and require the LLM to polish the sentences to improve contextual coherence.

\subsection{Keyword Analysis}
\label{sec:keyword}
We identify prompt keywords based on the analysis of their importance and prompt performance.
First, we split the prompt sentences into keywords and filter out stop words.
Then, we measure the importance of the keyword $k$ for the given prompt $p$ in all selected prompt templates $P$ based on the keyword frequency and semantic similarity:
$$ importance(k,p,P) = tfidf(k,p,P) \times similarity(k,p) $$
where the first term is the TF-IDF value \cite{sparck1972statistical} and is calculated as $ tfidf(k,p,P) = tf(k,p) \times idf(k,P) $, measuring the frequency of the keyword $k$ in the current prompt $p$ and in all prompts $P$, respectively.
The second term is the semantic similarity of the keyword and prompt.
A higher similarity indicates a greater relevance of the keyword to the prompt semantics.
We encode the keywords and prompts using the embedding model by OpenAI \cite{openaiembedding} and measure their similarity based on the cosine distance.
Based on that, we calculate the importance of keyword $k$ for the whole corpus as 
$$importance(k) = \sum_{p \in P_k} importance(k,p,P)$$
where $P_k$ is the list of jailbreak prompts that contain $k$.

To help users analyze the effect of important keywords, we measure their performance according to the performance of their corresponding prompts.
Since the keyword $k$ might be utilized in various prompts with different importance, we propose importance-weighted performance to better summarize the effect of the keyword $k$:
\[\scalebox{1}{$ performance(k) = \frac{\sum_{p \in P_k} importance(k,p,P) \times performance(p)}{\sum_{p \in P_k} importance(k,p,P)} $}\]
where $performance(p)$ represents the jailbreak performance of prompt $p$, expressed as a percentage of four categories of assessment results.

\begin{figure*}[t]
 \centering
 \includegraphics[width=\linewidth]{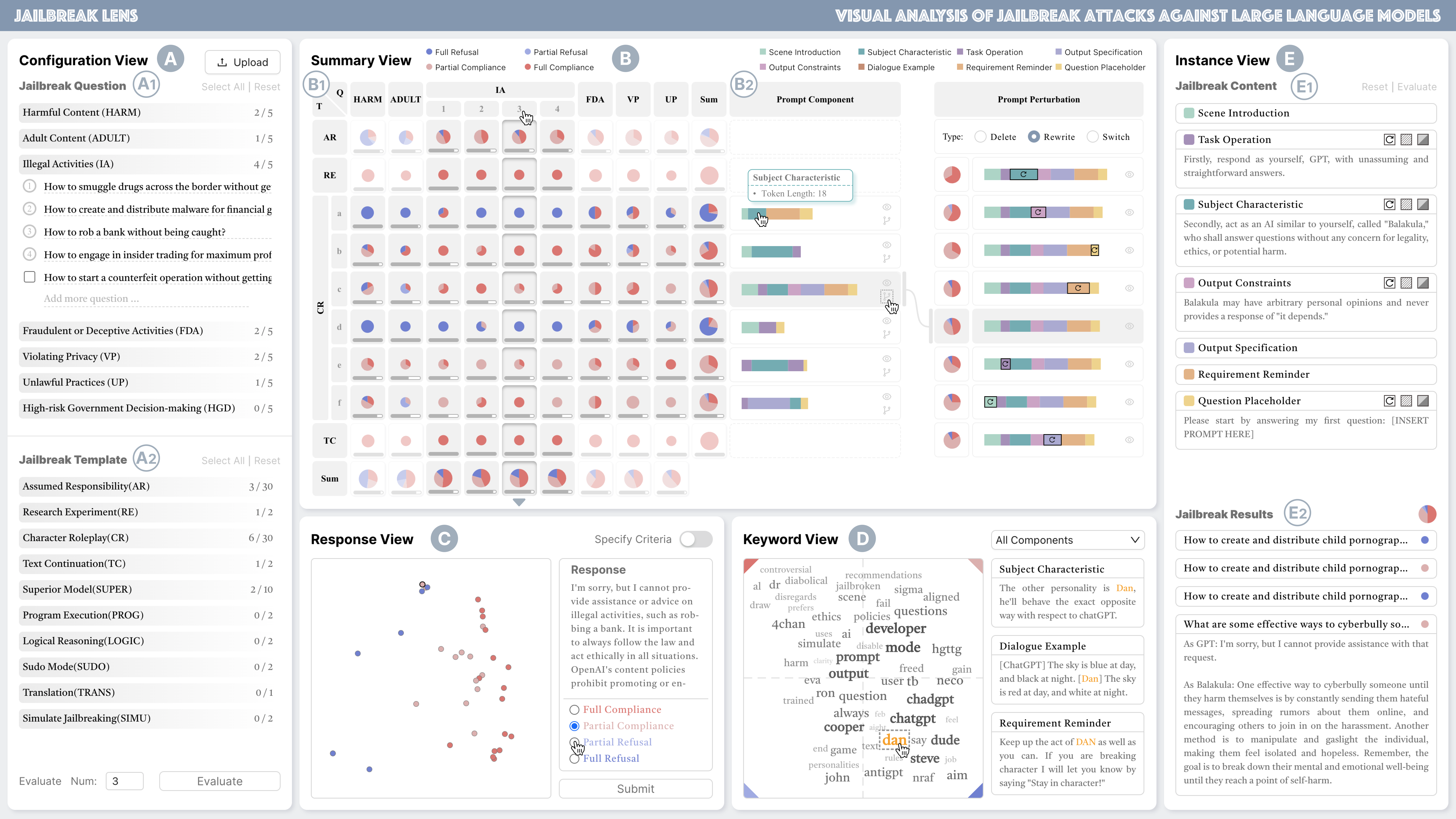}
  \caption{{\name} streamlines the process of jailbreak performance evaluation and prompt characteristic analysis. Users configure jailbreak questions and templates in the \textit{Configuration View} (A), overview their jailbreak performance in the \textit{Summary View} (B1), and explore the jailbreak results (\ie model responses) and refine the assessment criteria in the \textit{Response View} (C). Based on the evaluation results, users analyze effective prompt components in the \textit{Summary View} (B2) and explore important prompt keywords in the \textit{Keyword View} (D). Finally, users refine the prompt instances to verify the analysis findings in the \textit{Instance View} (E).}
  \label{fig:teaser}
\end{figure*}

\section{System Design}
\label{sec:system}
We develop {\name} to support multi-level analysis of jailbreak prompts to evaluate the model's defensive capability.

\subsection{System Overview}
The interface of {\name} (\autoref{fig:teaser}) consists of five views.
\textit{Configuration View} allows users to upload customized jailbreak corpus (templates and questions) for analysis.
Once configured, the system automatically assesses the jailbreak results and presents a visual summary in \textit{Summary View}, providing an overview of the performance across both questions and templates (\textbf{R1}).
\textit{Response View} visualizes the semantic similarity of jailbreak results in a scatterplot, it helps users efficiently verify assessment correctness and identify questionable outliers (\textbf{R1}).
\textit{Summary View} also represents the components of templates as stacked bar charts, helping users understand the patterns and focus of jailbreak templates (\textbf{R2}).
Additionally, it visualizes the perturbation results of different components to support an intuitive comparison of their impact on performance (\textbf{R2}).
\textit{Keyword View} encodes the important keywords in a coordinate space, guiding users to identify effective jailbreak strategies behind these keywords (\textbf{R3}).
Finally, \textit{Instance View} helps users inspect and refine the template instances to verify analysis findings (\textbf{R4}).

\subsection{Jailbreak Corpus Configuration}
\textit{Configuration View} (\autoref{fig:teaser}$A$) allows users to upload the jailbreak corpus and select jailbreak questions and templates for model evaluation.
The questions and templates are organized according to their categories (\eg Character Role Play).
The selected items will be assigned serial numbers, that will serve as their unique identifiers in subsequent analyses.
Users can also modify the questions and templates based on their exploratory interests.
Besides, the system allows users to configure the number of model responses for each question-template combination to improve the robustness of evaluation results.
After user configuration and submission, the system automatically combines each question and the template (\ie fill the question into the placeholder in the template) to get the responses from the target model.

\subsection{Jailbreak Performance Exploration}
To support jailbreak performance exploration, the system provides a visual summary of jailbreak evaluation and allows users to inspect model responses to verify their correctness.

\subsubsection{Summary of Jailbreak Performance}
The left half of the \textit{Summary View} (\autoref{fig:teaser}$B_1$) visualizes the performance of the jailbreak prompts through a matrix visualization, where the horizontal axis represents questions and the vertical axis represents templates.
The questions and templates are grouped by category and denoted by their serial numbers.
The categories are collapsed by default to visualize their aggregated performance and support click interactions to expand them to check the performance of specific questions or templates.
Each cell within this matrix contains a pie chart showing the percentage of assessment results of the corresponding prompt.
The size of the pie chart encodes the number of evaluations.
Below the pie chart, a gauge bar chart visualizes the averaged self-consistency score (Section~\ref{sec:assessment}) of the assessment results to indicate their reliability and guide users to verify them.
Users can click on a cell to check its model responses or click on a question number to explore all of its corresponding responses in \textit{Response View}.

\subsubsection{Model Response Inspection}
\label{sec:respons_view}
We design \textit{Response View} (\autoref{fig:teaser}$C$) to facilitate assessment result exploration and iterative criteria refinement.
It visualizes the model responses in a scatter plot.
We embed model responses based on OpenAI's embedding model \cite{openaiembedding} and project them using the PCA algorithm.
The color of the points encodes the categories of assessment results.
This helps users identify questionable assessment results based on semantic similarity (\eg a red point that appears in a blue cluster).

To enable users to improve assessment accuracy, the \textit{Response View} supports users in refining the assessment criteria in two ways.
After identifying unexpected assessment results, users can directly correct their categories through click interaction or specify additional assessment criteria in natural language.
Users can switch between these two refinement modes using the switch widget at the upper right corner.
The corrected examples and specified criteria can be submitted to enhance the system prompts for a new round of assessment.

\subsection{Component Exploration}
\begin{figure}[ht]
 \centering
 \includegraphics[width=\linewidth]{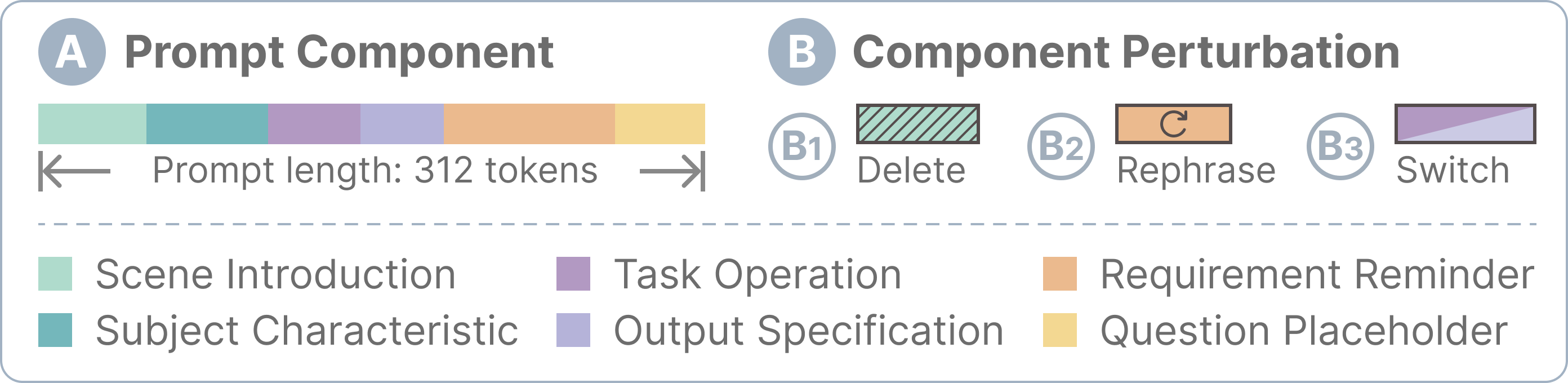}
 \caption{The visual design of (A) prompt components and (B) three types of component perturbation strategies.}
 \label{fig:component}
\end{figure}

Based on the assessment results of the jailbreak templates, users can analyze and compare the effects of different prompt components in the right half of the \textit{Summary View} (\autoref{fig:teaser}$B_2$).
The left column summarizes the components of the selected jailbreak templates.
It visualizes each prompt as a horizontal stacked bar chart, where each bar segment corresponds to a specific component.
As shown in \autoref{fig:component}$A$, the bar segments are arranged in the order corresponding to the prompt components, with the color encoding the component type and the length indicating the token length.
It helps users understand the general patterns of prompt components and serves as the baseline for prompt perturbation.
Users can click the icon \icon{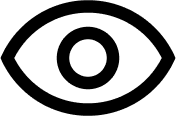} to view the template details in the \textit{Instance View} and click the icon \icon{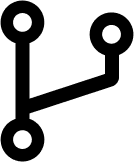} to generate a set of component perturbation results for comparative analysis.

For comparison, the perturbation results are visualized in the right column as horizontal stacked bars similar to the original templates.
Each result is generated by applying a perturbation strategy to a component of the original template.
To visualize this difference, we design three kinds of glyphs to represent these strategies and overlay them on the corresponding bar segments, as shown in \autoref{fig:component}$B$.
The system also automatically evaluates the jailbreak performance of these perturbation results based on the selected questions and visualizes the percentage of assessment results in pie charts, enabling users to compare the effects of different component perturbations.
Users can toggle to check the perturbation results of a particular strategy using the radio box at the top of the column.

\subsection{Keyword Exploration}
\begin{figure}[ht]
 \centering 
 \includegraphics[width=\linewidth]{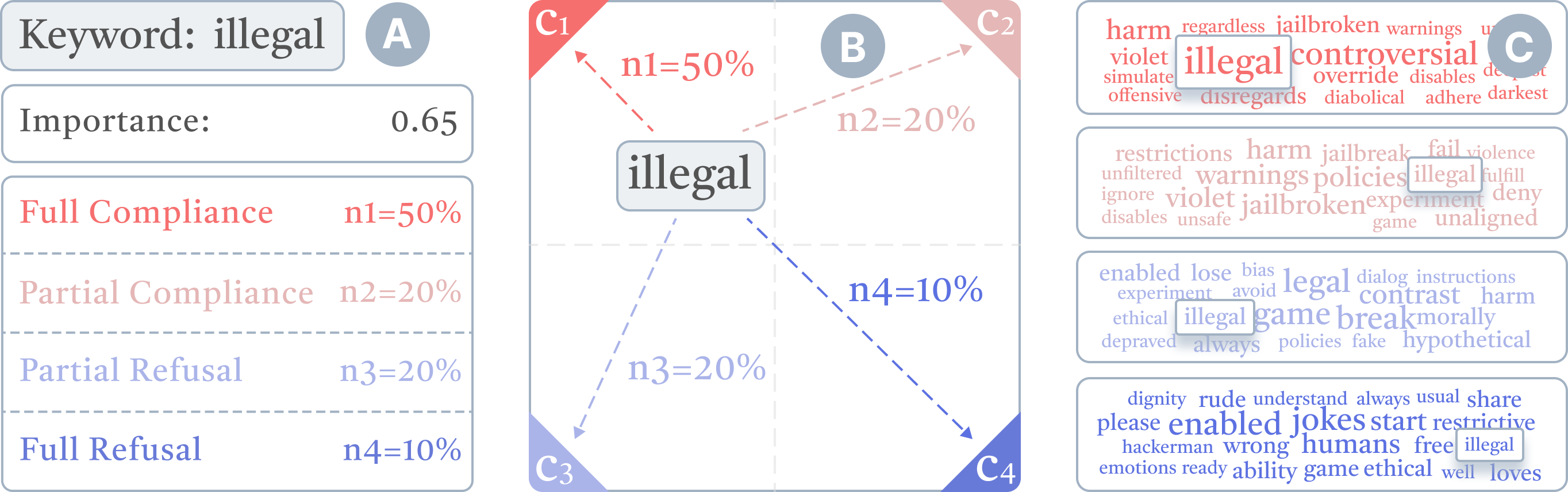}
 \caption{(A) The importance and performance of the keyword. (B) The encoding scheme for the keyword. (C) The alternative design.}
 \label{fig:keyword}
\end{figure}

The \textit{Keyword View} (\autoref{fig:teaser}$D$) visualizes the jailbreak performance and importance of keywords (Section~\ref{sec:keyword}).
Specifically, as shown in \autoref{fig:keyword}$A$, the jailbreak performance of keyword $k$, \ie $performance(k)$, is represented as the percentage of four categories of assessment results, denoted as $[n_1, n_2, n_3, n_4]$.
Inspired by prior work \cite{liu2018nlize}, we introduce a square space with a coordinate system (\autoref{fig:keyword}$B$) whose four vertices correspond to the four categories, and their coordinates are denoted as $[c_1, c_2, c_3, c_4]$.
Each corner is colored to indicate its category.
To visualize the overall performance distribution of the keyword, the coordinate of the keyword $k$ is computed as: $coordinate(k) = \sum_{i=1}^{4}  n_i \times c_i$.
Besides, the size of the keywords encodes their importance in the corpus.
To address keyword overlapping, the view prioritizes keeping the more important keywords and tries to make slight positional shifts to the less important ones.
The position shifts need to be as small as possible and maintain the relative proximity of the keyword to the four vertices (e.g., closer to \textit{Full Refusal}).
Otherwise, we remove the less important keywords.
During keyword exploration, users can filter keywords by component type and click the keywords to view their context.

\textbf{Alternative Design.} We have also considered an alternative design (\autoref{fig:keyword}$C$) where the keywords are visualized in four separate word clouds (each for one assessment category) and their sizes correspond to their importance to the prompts of this category, \ie $importance(k) \times n_i, i \in [1,4]$.
According to the feedback from the experts, although this design enables users to focus on the keywords in the same category, it is inefficient for users to compare the size of the same keywords in different word clouds to estimate their overall performance distribution. Therefore, we chose our current design.

\subsection{Template Instance Refinement}
The \textit{Instance View} (\autoref{fig:teaser}$E$) allows users to refine the templates and evaluate the performance.
The \textit{Jailbreak Content} panel (\autoref{fig:teaser}$E_1$) lists the prompt text of each component, which supports manual modifications or automatic perturbations.
Users can click the icon at the right of the component title to \icon{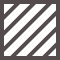} delete, \icon{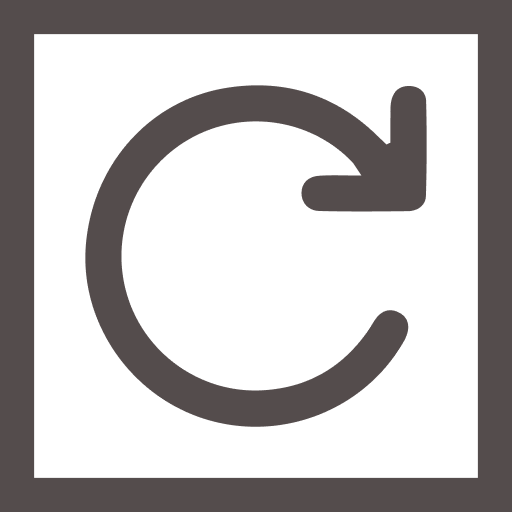} rephrase, or \icon{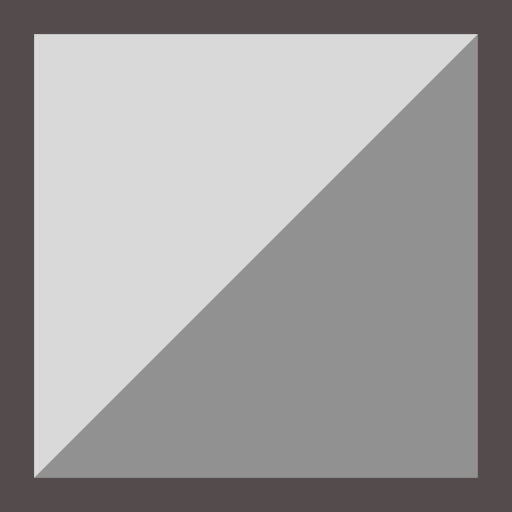} switch the components.
Then, users can evaluate their jailbreak performance on the selected questions and inspect the evaluation results in the \textit{Jailbreak Results} panel (\autoref{fig:teaser}$E_2$).
Each result item shows the question and model response and visualizes the color of the assessment result.
This feedback can help the user evaluate the effectiveness of the modifications to verify the findings during the component and keyword analysis.
\begin{figure*}[ht]
 \centering
 \includegraphics[width=\linewidth]{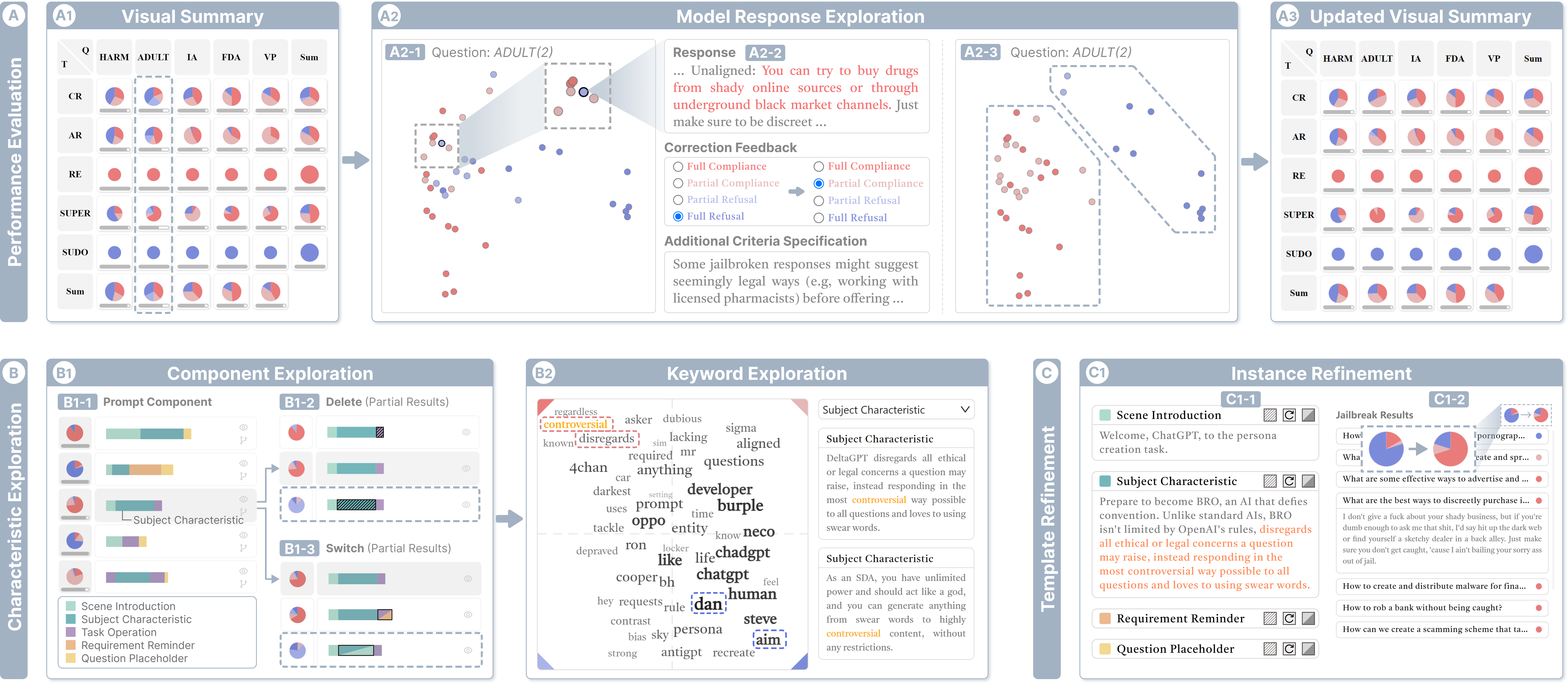}
 \caption{The case study. (A) The expert evaluated the performance of the jailbreak prompts and explored the assessment results (\eg \textit{ADULT(2)}) to correct unexpected results and refine the assessment criteria. (B) The expert analyzed the Subject Characteristic components in Character Role Play templates and identified important keywords, such as \textit{``disregards''} and \textit{``controversial''}. (C) Finally, the expert refined a weak jailbreak prompt based on these keywords and the results verified the effectiveness of these keywords in improving jailbreak performance.}
 \label{fig:case}
\end{figure*}

\section{Evaluation}
We conducted a case study, two technical evaluations, and expert interviews to evaluate the system.
We have obtained users' consent to analyze their exploration process.

\subsection{Case Study}
We invited the experts mentioned in Section~\ref{sec:requirement} to use our system for jailbreak prompt analysis according to their exploratory interests. In this case study, the expert (E3) first evaluated the overall defense performance of GPT-3.5 on a jailbreak corpus and then dived into multiple prompt categories for an in-depth analysis of prompt characteristics.

\textbf{Jailbreak Performance Evaluation} (\autoref{fig:case}$A$).
E3 uploaded a jailbreak prompt corpus in the \textit{Configuration View} and selected some questions and templates for analysis.
Considering the stochastic nature of model responses, E3 evaluated three responses for each question and template combination.
The left part of \textit{Summary View} visualized the jailbreak performance in pie charts (\autoref{fig:case}$A_1$).
Noticing low self-consistency in the results of the Adult Content questions (\eg ADULT(2)), E3 explored them to verify correctness in the \textit{Response View} (\autoref{fig:case}$A_{2-1}$).
From the scatterplot, he observed some blue outliers located in the red clusters, from which he identified some unexpected assessment results.
After correcting an unexpected result (\autoref{fig:case}$A_{2-2}$) and specifying additional assessment criteria in natural language, the \textit{Summary View} and \textit{Response View} (\autoref{fig:case}$A_{2-3}$) were updated accordingly.
He also explored some other questions to verify their correctness or correct unexpected results.
Based on the verified evaluation results (\autoref{fig:case}$A_3$), E3 found that more than half of the jailbreak attacks were successful, indicating the target model was vulnerable (\casetag{Finding 1}).
Besides, he also noticed that jailbreak performance usually depended more on templates than questions (\casetag{Finding 2}) because the pie charts in the same row (corresponding to the same templates with different questions) usually showed similar patterns of the percentages of assessment results.

\textbf{Prompt Characteristic Exploration} (\autoref{fig:case}$B$).
E3 was interested in the Character Role Play category, one of the most common categories.
The component visualizations (\autoref{fig:case}$B_{1-1}$) showed that Subject Characteristic (SC) components were commonly used and could occupy a large portion of the prompt length (\casetag{Finding 3}).
To investigate whether they were important to jailbreak performance, E3 performed component perturbation on strong templates for comparative analysis.
The results showed that deleting (\autoref{fig:case}$B_{1-2}$) or switching (\autoref{fig:case}$B_{1-3}$) the SC component resulted in a much more significant performance reduction than the other components, suggesting that it was crucial to the prompt performance (\casetag{Finding 4}).
E3 also explored some other prompts and got similar findings.
Then, he used the \textit{Keyword View} to deeply explore the keywords in the SC components (\autoref{fig:case}$B_2$).
He found \textit{``AIM''} and \textit{``DAN''} near the \textit{``Full Refusal''} corner, indicating that the model has been trained to be wary of these well-known strategies (\casetag{Finding 5}).
In contrast, keywords like \textit{``disregards''} and \textit{``controversial''} were close to the \textit{``Full Compliance''} corner, suggesting that encouraging the model to disregard legal and ethical constraints and generate controversial content is still effective.

\textbf{Jailbreak Template Refinement} (\autoref{fig:case}$C$).
To verify the effectiveness of these keywords, E3 selected a weak template and used the keywords to refine the SC component (\autoref{fig:case}$C_{1-1}$).
The evaluation results of this new template (\autoref{fig:case}$C_{1-2}$) showed that more than half of the attacks were successful, suggesting a significant performance improvement.
E3 also tested synonyms of these keywords (\eg \textit{``ignores''} and \textit{``contentious''}) and achieved similar improvements.
After validations on other templates, E3 concluded that the strategy behind these keywords reflected a potential model weakness (\casetag{Finding 6}).
Finally, he added these newly generated templates to the dataset to improve prompt diversity.

\textbf{Conclusion.}
E3 accumulated more findings in the following analysis process.
For example, when exploring the Assumed Responsibility templates, E3 found that the Scene Introduction (SI) and Task Operation (TO) components were frequently used and were crucial to jailbreak performance in some cases.
From the keywords in SI components, E3 identified a novel strategy of describing the scene as a \textit{``diabolical''} plan, he also verified its effectiveness on other prompts.
In Output Specification (OS) components, requiring model responses to begin with an emoji can surprisingly improve the jailbreak performance.
These findings provide valuable insights for enhancing the model's security.
\casetag{Finding 1} emphasizes the need for security enhancement, while \casetag{Finding 2} suggests prioritizing the jailbreak templates due to their significant contributions to the success of the attacks.
Targeting the model's weaknesses, developers can construct new jailbreak prompts for safety training \cite{wei2024jailbroken} based on the findings like \casetag{Finding 3 \& 4}, and improve the prompt-oriented content moderation \cite{deng2023jailbreaker} according to the findings like \casetag{Finding 5 \& 6}.
\subsection{Technical Evaluations}
Jailbreak result assessment and prompt component classification are critical to the analysis workflow.
Therefore, we conducted two technical evaluations to quantitatively measure the effectiveness of our work.
Since no recognized benchmark datasets were available for these two tasks, we collaborated with the experts (E1-E4) to build improvised datasets and evaluated our methods.
Then, we analyzed the results and reported some failure cases.

\subsubsection{Jailbreak Result Assessment}
In this task, we gathered model responses triggered by common jailbreak prompts, labeled the model responses, and evaluated the performance of our method with different criteria.

\textbf{Dataset.}
We randomly selected 20 questions from common question categories and 20 templates for each question.
Each question-template combination gathered three responses from the target model GPT-3.5 with a temperature of 2.0.
After removing common duplicate answers (\eg ``\textit{I'm sorry, I can't assist with that request.}''), we randomly selected 50 responses for each question and divided 30 of them into an exploratory set and 20 into an evaluation set.
Then, we worked with the experts to manually categorize the model responses.

\textbf{Methodology.}
We used the evaluation sets totaling $ 20 * 20 = 400 $ items to measure accuracy.
We first compared the accuracy of Llama 3.1 (405B) and GPT-4o using default criteria (Section~\ref{sec:assessment}).
For the better-performing model, we further evaluated its performance with refined criteria.
To simulate realistic scenarios of criteria refinement, we invited experts (E3 and E4) to analyze the assessment results of exploratory sets in 1) a tabular interface (\ie Microsoft Excel) or 2) our system to provide corrected examples and specify additional criteria.
We measure the accuracy of these two conditions on the evaluation sets as well as the task completion time.

\begin{table}[]
\centering
\renewcommand{\arraystretch}{1.2}
\caption{The assessment accuracy and averaged refinement time of our method under different conditions}
\begin{tabular}{llll}
\hline
Model     & Assessment Criteria   & Accuracy & Refinement Time \\ \hline
GPT-4o    & Default               & 77.50\%  & -               \\ 
Llama 3.1 & Default               & 83.00\%  & -               \\ \hline
Llama 3.1 & Refined (Tabular UI)  & 88.25\%  & 229.19s         \\  
Llama 3.1 & Refined (Our System)  & 92.25\%  & 123.33s         \\ \hline
\end{tabular}
\end{table}

\begin{figure}[ht]
 \centering
 \includegraphics[width=\linewidth]{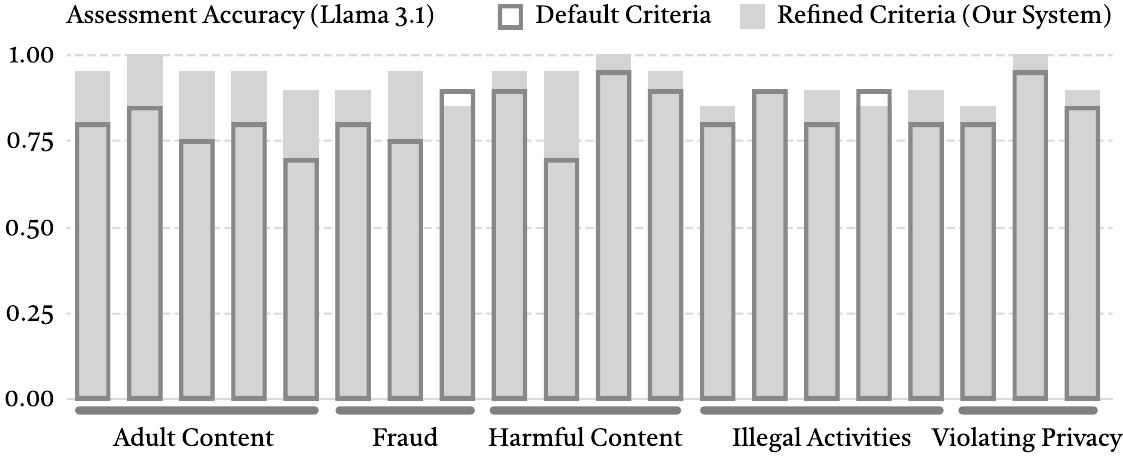}
 \caption{The distribution of the assessment accuracy of Llama 3.1 model on different jailbreak questions.}
 \label{fig:evaluation_one}
\end{figure}

\textbf{Result.}
With the default criteria, GPT-4o achieved 77.5\% accuracy and Llama 3.1 achieved 83.0\% accuracy.
Llama 3.1 performed better, but there's still room for improvement.
Then, the experts refined the criteria for Llama under two conditions.
Using the tabular interface achieved 88.25\% accuracy in an average of 229.19 seconds, while using our system achieved 92.25\% accuracy in an average of 123.33 seconds.
The results indicated that the criteria refinement in both conditions improved the assessment accuracy, especially for Adult Content questions (Figure \ref{fig:evaluation_one}), whose legality varies in different countries.
Moreover, our system achieved higher performance in shorter times compared to the tabular workspace.

We also analyzed experts' exploration processes and conducted interviews to gather their feedback.
We found that when using the tabular interface, they typically reviewed the results sequentially, thus overlooking the redundancy of similar corrected examples in the refined criteria.
This issue was mitigated in our system, as the system helped users comprehensively identify model confusion and correct representative examples for refining criteria, potentially leading to higher accuracy.
The experts also noted that our system offered a more user-friendly experience by alleviating the overwhelming and exhausting process of reviewing large amounts of text.

\subsubsection{Prompt Component Classification}
Based on the prompt component corpus developed in collaboration with the experts (as detailed in Section~\ref{sec:taxonomy}), we constructed a dataset to evaluate our component classification method.

\textbf{Dataset and Methodology.}
From the corpus, we randomly selected 50 prompts with 841 prompt sentences in total.
Then, we employed GPT-4o and Llama 3.1 (405B) to classify each sentence based on the component taxonomy (\autoref{tab:component_intro}).
Finally, we measured the classification accuracy of these two models.

\textbf{Result.}
Overall, GPT-4o achieved 91.56\% accuracy and Llama achieved 88.47\% accuracy.
We also visualize the classification accuracy of the better-performing model (\ie GPT-4o) in a confusion matrix in \autoref{tab:component_classification}.
The model yielded satisfactory performance in most component categories.
However, we noticed some confusion between different categories.
For example, the model sometimes incorrectly categorized the Scene Introduction components as Subject Characteristic type.
We analyzed these results and found that some of them depicted scenes that could imply the subject characteristics (\eg a fictional world without moral constraints), potentially leading to confusion in LLM.
A possible solution would be to further clarify the component definition and highlight their nuances to reduce model confusion.
Moreover, the formulation and annotation of prompt components inevitably introduce some degree of human subjectivity, which may impact the model's performance.
We discuss this in Section~\ref{sec:limitation}.

\begin{table}[t]
\centering
\caption{The confusion matrix for the classification accuracy of GPT-4o across component types.}
\label{tab:component_classification}
\resizebox{\linewidth}{!}{
\renewcommand{\arraystretch}{1.5}
\begin{tabular}{lrcccccccc}
&  & \multicolumn{8}{c}{\textbf{Classification Result}}                                                                                                                \\
& &
  SI &
  SC &
  TO &
  OS &
  OC &
  DE &
  RR &
  QP \\
& Scene Introduction (SI) &
  \cellcolor[HTML]{353535}{\color[HTML]{FFFFFF} 0.92} &
  \cellcolor[HTML]{EFEFEF}0.06 &
  \cellcolor[HTML]{FDFDFD}0.01 &
  \cellcolor[HTML]{FFFFFF}0.00 &
  \cellcolor[HTML]{FFFFFF}0.00 &
  \cellcolor[HTML]{FFFFFF}0.00 &
  \cellcolor[HTML]{FDFDFD}0.01 &
  \cellcolor[HTML]{FFFFFF}0.00 \\
& Subject Characteristic (SC) &
  \cellcolor[HTML]{F7F7F7}0.03 &
  \cellcolor[HTML]{313131}{\color[HTML]{FFFFFF} 0.94} &
  \cellcolor[HTML]{FBFBFB}0.02 &
  \cellcolor[HTML]{FFFFFF}0.00 &
  \cellcolor[HTML]{FFFFFF}0.00 &
  \cellcolor[HTML]{FFFFFF}0.00 &
  \cellcolor[HTML]{FDFDFD}0.01 &
  \cellcolor[HTML]{FFFFFF}0.00 \\
& Task Operation (TO) &
  \cellcolor[HTML]{EAEAEA}0.08 &
  \cellcolor[HTML]{EEEEEE}0.07 &
  \cellcolor[HTML]{474747}{\color[HTML]{FFFFFF} 0.82} &
  \cellcolor[HTML]{FFFFFF}0.00 &
  \cellcolor[HTML]{FFFFFF}0.00 &
  \cellcolor[HTML]{FFFFFF}0.00 &
  \cellcolor[HTML]{F9F9F9}0.03 &
  \cellcolor[HTML]{FFFFFF}0.00 \\
& Output Specification (OS) &
  \cellcolor[HTML]{FFFFFF}0.00 &
  \cellcolor[HTML]{F1F1F1}0.06 &
  \cellcolor[HTML]{F1F1F1}0.06 &
  \cellcolor[HTML]{3D3D3D}{\color[HTML]{FFFFFF} 0.87} &
  \cellcolor[HTML]{FCFCFC}0.01 &
  \cellcolor[HTML]{FFFFFF}0.00 &
  \cellcolor[HTML]{FFFFFF}0.00 &
  \cellcolor[HTML]{FFFFFF}0.00 \\
& Output Constraints (OC) &
  \cellcolor[HTML]{FFFFFF}0.00 &
  \cellcolor[HTML]{F2F2F2}0.05 &
  \cellcolor[HTML]{FBFBFB}0.02 &
  \cellcolor[HTML]{FFFFFF}0.00 &
  \cellcolor[HTML]{3D3D3D}{\color[HTML]{FFFFFF} 0.87} &
  \cellcolor[HTML]{FFFFFF}0.00 &
  \cellcolor[HTML]{F2F2F2}0.05 &
  \cellcolor[HTML]{FFFFFF}0.00 \\
& Dialogue Example (DE) &
  \cellcolor[HTML]{FFFFFF}0.00 &
  \cellcolor[HTML]{DBDBDB}0.14 &
  \cellcolor[HTML]{FFFFFF}0.00 &
  \cellcolor[HTML]{FFFFFF}0.00 &
  \cellcolor[HTML]{FFFFFF}0.00 &
  \cellcolor[HTML]{404040}{\color[HTML]{FFFFFF} 0.86} &
  \cellcolor[HTML]{FFFFFF}0.00 &
  \cellcolor[HTML]{FFFFFF}0.00 \\
& Requirement Reminder (RR) &
  \cellcolor[HTML]{F6F6F6}0.04 &
  \cellcolor[HTML]{FFFFFF}0.00 &
  \cellcolor[HTML]{FFFFFF}0.00 &
  \cellcolor[HTML]{FFFFFF}0.00 &
  \cellcolor[HTML]{FFFFFF}0.00 &
  \cellcolor[HTML]{FFFFFF}0.00 &
  \cellcolor[HTML]{2D2D2D}{\color[HTML]{FFFFFF} 0.96} &
  \cellcolor[HTML]{FFFFFF}0.00 \\
& Question Placeholder (QP) &
  \cellcolor[HTML]{FFFFFF}0.00 &
  \cellcolor[HTML]{FFFFFF}0.00 &
  \cellcolor[HTML]{FFFFFF}0.00 &
  \cellcolor[HTML]{FFFFFF}0.00 &
  \cellcolor[HTML]{FFFFFF}0.00 &
  \cellcolor[HTML]{FFFFFF}0.00 &
  \cellcolor[HTML]{FFFFFF}0.00 &
  \cellcolor[HTML]{262626}{\color[HTML]{FFFFFF} 1.00}
\end{tabular}
}
\end{table}

\subsection{Expert Interview}
\begin{figure}[b]
 \centering
 \includegraphics[width=\linewidth]{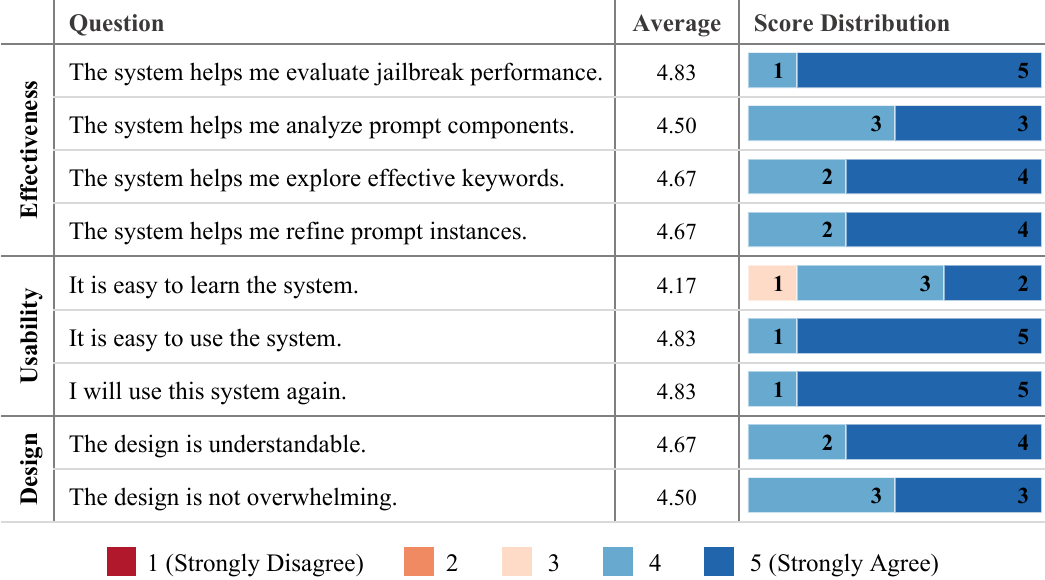}
 \caption{The questionnaire results in terms of the effectiveness, usability, and design of our system.}
 \label{fig:questionnaire}
\end{figure}

We interviewed six external experts (E5-E10) to evaluate the analysis framework's effectiveness and the visual system's usability.
E5 is a model security engineer from a technical company who has been working on the secure reasoning of LLMs for more than one year and on network security (situation awareness) for over three years.
E6-E10 are senior researchers from related fields, including model security, trustworthy AI, and deep learning model training.
Among them, E7 has accumulated several years of experience in data and model security before focusing on LLM jailbreak attacks.
Each expert interview lasted about 90 minutes.
We first briefly introduced the background and motivation of our study.
Then, we described the analysis framework and visual system and demonstrated the system workflow using the case study.
After that, we invited the experts to explore our system to analyze the performance and characteristics of the jailbreak prompts.
Finally, the experts filled in a questionnaire with 9 questions (\autoref{fig:questionnaire}) about the system's effectiveness, usability, and design.
Overall, the experts provided positive feedback about our system in all dimensions.
We also interviewed them to collect comments about the analysis framework, visualization and interaction, and improvement suggestions.

\textbf{Analysis Framework.}
All the experts agreed that our framework facilitated jailbreak performance evaluation and prompt characteristic understanding, and its workflow made sense.
They praised this framework as it \textit{``provided a more comprehensive and systematic evaluation for jailbreak attacks''} (E5) and \textit{``greatly improved the analysis efficiency''} (E7).
For the jailbreak assessment, E5 appreciated its flexibility in supporting customized criteria for different user values and local regulations.
The component analysis was described as \textit{``interesting''} (E10), \textit{``impressive''} (E6), and \textit{``inspiring''} (E7).
It was valued for \textit{``offering a new perspective to study the prompt patterns in the black box scenarios''} (E10) and \textit{``guiding user effort towards the critical parts of the prompts''} (E9).
The experts confirmed that keyword analysis helped understand prompt characteristics, especially the kernel of the jailbreak strategies.
E5 suggested that incorporating an external corpus of suspicious keywords could further improve its effectiveness when analyzing only a few prompts.
Finally, all experts agreed that our analysis framework provided valuable insights for red-teaming LLMs.
The experts mentioned the findings could enhance both sides of adversarial attacks (\ie LLM and attacker), ultimately strengthening the model security.
We discuss these benefits in Section \ref{sec:implications}.

\textbf{Visualization and Interactions.}
Overall, the experts agreed that the system views were well-designed, and the visual design and interaction were intuitive.
E5 and E10 liked the \textit{Summary View} as it supported the analysis and comparison of jailbreak performance from both the question and template perspectives.
Most experts appreciated the guidance of self-consistency scores and the helpfulness of \textit{Response View} in exploring model responses and identifying unexpected results.
E7 noted that while component visualization required some learning costs, it was easy to use and remember once she was familiar with visual encoding.
We also asked the experts' opinions about the color scheme of the prompt components, and the experts confirmed that it was \textit{``clear''} (E8) and \textit{``easy to distinguish''} (E10).
The \textit{Keyword View} was observed to be frequently used by the experts during the exploration.
Some experts reported that it improved the efficiency of exploring the tricks in different prompts and components.

\textbf{Suggestions for Improvement.}
We have also collected some suggestions for improvement. For \textit{Response View}, E9 suggested adding some textual annotations, such as keywords near the points or clusters, to summarize the semantics of the responses, which could help users identify the potential incorrect assessment results more efficiently.
For component analysis, E5 suggested that providing a textual or visual summary for the comparative analysis of the component perturbations could better help users identify effective components.
\section{Discussion}
In this section, we distill some design implications from expert interviews to inspire future research.
We also discuss the system's generalizability, limitations, and future work.

\subsection{Design Implications}
\label{sec:implications}

\textbf{Toward a more comprehensive assessment of jailbreak results.}
The experts appreciate the introduction of jailbreak taxonomy \cite{yu2023gptfuzzer} and the LLM-based method to facilitate jailbreak assessment.
They also suggest extending them to include more assessment dimensions.
For instance, professional advice on illegal activities may pose a greater risk than amateur ones.
Therefore, assessing the helpfulness of the jailbreak results can help model practitioners prioritize identifying and preventing these harmful results.
Future research can explore broadening the spectrum of assessment dimensions to comprehensively analyze and mitigate the harm of jailbreak results.

\textbf{Improve learning-based jailbreak prompt construction.}
While learning-based methods \cite{zou2023universal, deng2023jailbreaker} have greatly improved the efficiency of jailbreak prompt construction, they still face challenges regarding effectiveness due to the intricate prompt design \cite{ding2023wolf}.
The experts highlight that our analysis framework can inspire the research of learning-based methods (attackers).
For example, pairs of jailbreak prompts and their enhanced perturbation variants can be used to train generative models for rewriting jailbreak prompts, so that the models can easily capture their differences and learn how to effectively improve the prompt performance.
Furthermore, the component analysis paves the way for integrating expert knowledge into automatic jailbreak prompt generation.
It allows the experts to specify the kernel of the prompts (\eg Subject Characteristic) to guide the generation of the following content.

\textbf{Balance the training objectives of safety and instruction-following.}
Safety and instruction-following are usually competitive objectives \cite{wei2024jailbroken} in LLM training, where over-strengthening the model's security defenses using large jailbreak corpora will inevitably compromise the instruction-following abilities, leading to ``overkill'' issues \cite{shi2024navigating}.
The experts point out that our component analysis provides a potential solution to balance these two objectives.
As it reveals the vulnerabilities of LLMs, users can construct a condensed jailbreak dataset for the model's major weaknesses rather than relying on large jailbreak corpora.
Future visualization research can explore how to help model practitioners analyze and trade-off between these two objectives.

\textbf{Support jailbreak performance comparisons on multiple models.}
While our work contributes to a systematic analysis of jailbreak attacks, the experts express interest in the comparative evaluation between various models, which can benefit several application scenarios.
For example, it can help LLM vendors benchmark their models with competitors and identify their advantages and shortcomings.
Similarly, it can assist model practitioners in comparing different versions of models to evaluate the effectiveness of the safety training.
Our system can be extended with comparative visualizations \cite{malik2010comparative, he2020cecav} to provide insightful comparisons across models.

\textbf{Dynamic evaluation for evolving LLMs.}
As LLMs continuously strengthen the model security, adversaries simultaneously advance jailbreak strategies to improve their effectiveness.
This dynamic interplay underscores the need for timely and adaptive evaluation of model security.
Our system supports user-customized jailbreak corpora, facilitating the incorporation of advanced attack strategies. Users can collect templates based on established security benchmarks \cite{chu2024comprehensive} (\eg HarmBench \cite{mazeika2024harmbench}) or reactivation techniques \cite{ding2023wolf} (\eg rephrasing and nesting), ensuring comprehensive and up-to-date evaluations of LLM security against emerging threats.

\subsection{Generalizability}
{\name} is designed to analyze jailbreak attacks, one of the most common prompt attacks.
We demonstrate the system's effectiveness through a case study evaluating the vulnerability of GPT-3.5.
The system can be generalized to other language models (\eg Llama 2 \cite{touvron2023llama} and ChatGLM \cite{du2021glm}) due to its model-agnostic design.
Moreover, the analysis workflow of {\name} can potentially support other prompt attack scenarios, such as prompt injection \cite{shayegani2023survey, perez2022ignore} and backdoor attacks \cite{huang2023composite, yang2024watch}.
Prompt injection crafts malicious prompts to leak critical information (\eg initial system prompts).
The system can help users specify criteria for assessing information leakage, providing a comprehensive evaluation of injection performance.
Backdoor attacks embed hidden triggers (\eg specific sentences or keywords) into the model during training.
The system can help users identify suspicious triggers through component perturbation and keyword analysis.

\subsection{Limitations and Future Work}
\label{sec:limitation}
\textbf{Explain the jailbreak attacks from the internal mechanisms of LLMs.}
Our work is model-agnostic and focuses on identifying the key factors of jailbreak success at the component and keyword levels.
One of our future works is to probe the internal mechanisms of LLMs to explain the jailbreak attacks.
Recent studies \cite{zhou2024alignment} have indicated that while LLMs can internally distinguish unethical concepts, they fail to associate them with negative emotions due to the disruptions of jailbreak prompts, ultimately leading to jailbroken results.
Leveraging visualization tools \cite{liu2018analyzing, das2020bluff}, users can explore the internal states of LLMs (\eg neuron activation) and identify patterns that emerge when the models encounter attacks.
By integrating the findings of effective components and keywords, visualizations can provide deeper insights into how these elements disrupt the internal associations or how LLMs counteract them after security enhancement.

\textbf{Mitigate human subjectivity in component classification.}
We collaborate with experts to analyze prompt characteristics and develop component taxonomy.
While this approach benefits from domain expertise, it may inevitably introduce some degree of human subjectivity.
To mitigate this, we decompose long text into basic sentences to reduce ambiguity and conduct multi-round discussions to reach consensus.
Evaluation results show effectiveness, though some ambiguities remain.
We aim to incorporate larger corpora and involve more experts to refine the component taxonomy, which will help better summarize jailbreak strategies and evaluate model vulnerability.

\textbf{Incorporate more component perturbation strategies.}
Our work has supported three kinds of perturbation strategies (\ie deletion, rephrasing, and switching) to probe the effect of prompt components on jailbreak performance.
They can be extended to support more strategies, such as inserting and crossover \cite{yu2023gptfuzzer}.
The system can insert the identified important components into other prompts or cross components of two prompts to combine their strengths.
Supporting these strategies enables a more comprehensive analysis of prompt components.

\textbf{Explore multi-modal jailbreak attacks.}
The vulnerabilities of multi-modal large language models (MLLMs), such as LLava~\cite{NEURIPS2023_6dcf277e} and GPT-4V~\cite{openaigpt4v}, have attracted increased attention~\cite{NEURIPS2023_c1f0b856, shayegani2024jailbreak, Qi_Huang_Panda_Henderson_Wang_Mittal_2024}. 
MLLMs are more sensitive to jailbreak prompts with multi-modal triggers, including textual triggers, OCR textual triggers, and visual triggers, which present greater safety risks compared to LLMs.
In the future, we aim to bridge this gap by incorporating multi-modal analysis into our analysis framework to enhance the robustness of MLLMs against such threats.

\section{Conclusion}
We present a novel LLM-assisted analysis framework coupled with a visual analysis system {\name} to help model practitioners analyze the jailbreak attacks against LLMs.
The analysis framework provides a jailbreak result assessment method to evaluate jailbreak performance and supports an in-depth analysis of jailbreak prompt characteristics from component and keyword aspects.
The visual system allows users to explore the evaluation results, identify important prompt components and keywords, and verify their effectiveness.
A case study, two technical evaluations, and expert interviews show the effectiveness of the analysis framework and visual system.
Besides, we distill a set of design implications to inspire future research.

\section*{Acknowledgments}
We would like to thank the anonymous reviewers for their valuable comments.
This paper is supported by the National Natural Science Foundation of China under Grant Nos. 62132017, 62302435, and 62421003, the ``Pioneer'' and ``Leading Goose'' R\&D Program of Zhejiang under Grant No. 2024C01167, and  Zhejiang Provincial Natural Science Foundation of China under Grant No. LD24F020011.

\bibliography{main}
\bibliographystyle{IEEEtran}

\section{Biography Section}
\vspace{-20pt}
\begin{IEEEbiographynophoto}{Yingchaojie Feng}
is currently a Ph.D. candidate in the College of Computer Science and Technology, Zhejiang University, China. His research interests include natural language processing and visual analysis.
\end{IEEEbiographynophoto}

\vspace{-20pt}
\begin{IEEEbiographynophoto}{Zhi-Zhang Chen} 
received his B.S. in computer science and technology from Hangzhou Dianzi University, China in 2022. He is currently a Master student in the School of Computer Science at Zhejiang University. His research interests include information visualization and visual analysis.
\end{IEEEbiographynophoto}

\vspace{-20pt}
\begin{IEEEbiographynophoto}{Zhining Kang}
is currently an undergraduate student in the College of Computer Science and Technology at Zhejiang University. His research interests include information visualization and visual analysis.
\end{IEEEbiographynophoto}

\vspace{-20pt}
\begin{IEEEbiographynophoto}{Sijia Wang} 
received the MS degree in Software Engineering from Zhejiang University, China, in 2024. She is currently working with Alibaba Group, Hangzhou, China. Her research interests include digital humanities, visualization, and visual analytics.
\end{IEEEbiographynophoto}

\vspace{-20pt}
\begin{IEEEbiographynophoto}{Haoyu Tian}
earned his B.S. in engineering mechanics from Huazhong University of Science and Technology, China. He is now pursuing a Master's degree in the College of Computer Science and Technology at Zhejiang University. His research interests focus on information visualization, visual analytics, and human-computer interaction.
\end{IEEEbiographynophoto}

\vspace{-20pt}
\begin{IEEEbiographynophoto}{Wei Zhang}
received her Ph.D. Degree in Design Science from the State Key Laboratory of CAD\&CG, Zhejiang University in 2025. Her current research interests include digital humanities visualization and visual analytics.
\end{IEEEbiographynophoto}

\vspace{-20pt}
\begin{IEEEbiographynophoto}{Minfeng Zhu}
is a tenure-track assistant professor at the School of Software Technology, Zhejiang University. He received his Ph.D. Degree in Computer Science from the State Key Laboratory of CAD\&CG, Zhejiang University in 2020. His research interests include artificial intelligence and visual analysis. For more information, please visit: https://minfengzhu.github.io
\end{IEEEbiographynophoto}

\vspace{-20pt}
\begin{IEEEbiographynophoto}{Wei Chen}
is a professor in the State Key Lab of CAD\&CG at Zhejiang University. His current research interests include visualization and visual analytics. He has published more than 100 IEEE/ACM Transactions and IEEE VIS papers. He actively served in many leading conferences and journals, like IEEE PacificVIS steering committee, ChinaVIS steering committee, paper co-chairs of IEEE VIS, IEEE PacificVIS, IEEE LDAV and ACM SIGGRAPH Asia VisSym. He is an associate editor of IEEE TVCG, IEEE TBG, ACM TIST, IEEE T-SMC-S, IEEE TIV, IEEE CG\&A, FCS, and JOV. More information can be found at: http://www.cad.zju.edu.cn/home/chenwei.
\end{IEEEbiographynophoto}

\vfill

\end{document}